\documentclass[11pt]{article} 
\pdfoutput=1 

\usepackage{jheppub} 

\usepackage[utf8]{inputenc}
\usepackage{ulem}
\usepackage{slashed}
\usepackage[english]{babel}
\usepackage{amsmath,amssymb,amsbsy,amstext,amsthm,booktabs,simplewick,exscale,relsize,slashed,graphicx,amsfonts,upgreek, xcolor,subcaption}
\usepackage{multirow,color,dcolumn,bm,enumerate}
\usepackage{hyperref}
\usepackage{dsfont}
\usepackage{ragged2e} 
\DeclareUnicodeCharacter{2212}{-}

\usepackage{hyperref}

\title{
Phenomenology of Fractionally Charged Particles: \\ Two Reps Are Better Than One}
\author[]{Seth Koren \&}
\emailAdd{skoren@nd.edu}
\author[]{Adam Martin}
\emailAdd{amarti41@nd.edu}
\affiliation[]{Department of Physics and Astronomy, University of Notre Dame, South Bend, IN, 46556 USA}

\abstract{
We continue our study of fractionally charged particles (FCPs)---particles carrying electric charge a multiple of $e/6$. Discovering an FCP would inform us about both Standard Model physics (what the true gauge group and the one-form global symmetry are) and Beyond the Standard Model physics (ruling out many unified theories), which makes them a high-stakes target for collider searches. Here we find that with two FCPs there are vastly richer phenomenologies compared to the single-particle extensions we previously studied.
Stringent constraints on colored FCPs can be dramatically weakened when decays are open; conversely the cross sections of the least visible species can be enlarged by up to $\sim 10^3$, increasing their discovery potential enormously at the LHC and milliQan. Overall, these simple models motivate performing searches for FCPs produced along with jets or leptons, and highlight `free' discovery potential in  reanalyzing existing missing-energy datasets for low-quality tracks.
}

\begin{document}
\maketitle

\setcounter{page}{2}

\section{Introduction}

\subsection{Theoretical background}

What is the fundamental quantum of charge in QED? With the known Standard Model species (and despite beginning with particles having hypercharge in multiples of $1/6$) we end up in infrared QED with every color singlet having an electric charge that is an integer multiple of $e$, the charge of the electron.  Here we briefly remind the reader of the theoretical structure behind the existence of fractionally charged particles (FCPs)---particles carrying electric charge a multiple of $e/6$---referring to our previous \cite{Koren:2024xof} for further details.

This quantization could be a curious coincidence, or it could be a hint of grand unification, as minimal models of gauge coupling unification in fact dictate this feature of infrared physics. Technically this is due to the fact that the Standard Model gauge group which emerges from those GUTs has nontrivial `global structure'
\begin{align}
    &G_{\rm SM_n} = \left(SU(3)_C \times SU(2)_L \times U(1)_Y\right)/\mathbb{Z}_n, \qquad n = 1,2,3,6, 
\end{align}
where we can have a nontrivial quotient group by a diagonal element of the centers of hypercharge and the non-Abelian groups \cite{Hucks:1990nw}. These different gauge \textit{groups} all share the same \textit{Lie algebra}, which is all that we have experimentally determined. In $G_{\rm SM_n}$, the fundamental quantum of charge in QED is $Q_{\rm min} = n e/6$ \cite{Tong:2017oea,Koren:2024xof,Alonso:2024pmq,Li:2024nuo,Alonso:2025rkk} (the possible axion couplings also depend on $n$, see \cite{Tong:2017oea,Reece:2023iqn,Choi:2023pdp,Cordova:2023her}). In $SU(5)$ or $SO(10)$ unification, the full $n=6$ quotient is demanded, while Pati-Salam demands $n=3$ and trinification demands $n=2$ (at the least; one can also consider versions of these with $n=6$). Therefore the discovery of an FCP would rule out some or all of the simplest models of unification.

Grand unification has long been held as an inevitability for ultraviolet physics, but its predictions (e.g. proton decay, or less directly weak scale supersymmetry) have not yet been borne out in data. It is increasingly prudent for us as a field to question our assumptions about what physics looks like Beyond the Standard Model, so it is remarkable that there are collider signatures accessible at the Large Hadron Collider that could \textit{falsify grand unification!} Aside from any ultraviolet motivation, there is also the simple fact that this structural ambiguity prevents us from actually knowing which `Standard' Model we have. Clearly these signatures are important targets for collider searches.

Different choices for the global structure of the gauge group preserve the local, perturbative physics, but modify nonperturbative, topological aspects. As stated, our focus is on the modification to representation theory: the quotient by $\mathbb{Z}_n$ sets a diagonal combination of gauge group generators equal to identity transformation, implying that fields are not allowed to transform non-trivially under that combination of generators. This demands (some of) the correlations between charges which are seen among the Standard Model fields to be a requirement of the gauge theory rather than an accident. See Section 6 of \cite{Koren:2024xof} for more pedagogical discussion of the representation theory of the different Standard Models. 

Since gauge symmetries are really mere redundancies, it is also useful to know that there is a gauge-invariant description of the different possibilities in terms of the spectrum of generalized global symmetries they possess. In particular, the SM with quotient $\mathbb{Z}_n$ has a $\mathbb{Z}_{6/n}^{(1)}$ electric one-form symmetry. Physically this means that the electric fields of heavy probe particles with fractional electric charge $n e/6$ cannot be screened by the Standard Model species. See Section 7 of \cite{Koren:2024xof} for more pedagogical discussion of the electric one-form symmetry of the Standard Models. Various uses of higher-form symmetries in particle physics have gradually been understood in recent years, e.g.  \cite{Cordova:2022qtz,Cordova:2022fhg,Cordova:2024ypu,
Delgado:2024pcv,
Brennan:2020ehu,Cordova:2022ieu,Cordova:2023her,Brennan:2023kpw, 
Anber:2021upc,Yokokura:2022alv,
Choi:2022jqy,Choi:2022fgx,Choi:2023pdp,
Reece:2023iqn,Aloni:2024jpb,
Putrov:2023jqi,
Craig:2024dnl}.

\subsection{Experimental background}

Discovering an FCP at a collider would tell us more than perhaps any other discovery about both the Standard Model and about ultraviolet physics, so it is clear that a robust experimental program is necessary. However, as we highlighted in \cite{Koren:2024xof}, there has been relatively little work in this direction in recent years. One notable exception is a CMS search \cite{CMS:2024eyx} for anomalously low $dE/dx$ in tracker hits---however, this search loses sensitivity precipitously for $Q/e \lesssim 1/2$ because of its dependence on tracker information for triggering, since tracks of low-charged particles are less likely to be reconstructed.\footnote{The CMS search \cite{CMS:2024eyx} remarks that below this there is an `increased probability' of tracks not being reconstructed. The related thesis \cite{Vannerom:2693948} only shows the $Q/e = 1/3, 2/3, 1$ benchmarks (Figure 6.7). For $2/3$ the efficiency is still nearly $1$, while for $1/3$ it's nearly $0$. A fuller understanding of the detector response over this range is important information to guide search proposals.} Meanwhile for $Q/e > 2/3$, it becomes difficult to tell these apart from charge 1, meaning among possible FCPs there is good coverage mostly only for $Q/e = 1/2$ or $2/3$.

Another important, dedicated effort is from milliQan with their demonstrator `bar detector' \cite{Ball:2020dnx} and ongoing upgrades including a new `slab detector' \cite{milliQan:2021lne}. milliQan will fill an important hole in coverage for $Q/e = 1/6$, but their search will be limited because of their small angular acceptance.
These ongoing experimental efforts are great, but needless to say there does not yet exist a comprehensive program to search for FCPs, so there is enormous room for innovation in search and analysis methods. These and other searches for millicharged particles can be sensitive to FCPs, but we emphasize again here that the existence of our models relies only on the SM structure, and they do not include a dark photon \cite{Galison:1983pa,Holdom:1985ag,Holdom:1986eq,Foot:1992ui}. Since the SM structure restricts their charge to multiples of $e/6$---quite large in terms of millicharged particle phenomenology---colliders should be able to offer generally better reach, if such searches are devised. 

The appearance of FCPs in certain models has long been understood \cite{Li:1981un,Goldberg:1981jt,Dong:1982sa,Frampton:1982gc,Kang:1981nr,Wen:1985qj,Schellekens:1989qb}, but this had been seen as a bug, rather than a feature \cite{Langacker:2011db}. A theoretical bias which assumed high reheating temperature would mean that these particles would necessarily have been present in the early universe plasma, leading to a relic abundance which was inconsistent with observations. However, in recent times this assumption has been under question as CMB measurements have placed lower and lower upper bounds on the scale of inflation \cite{Planck:2018jri}. There has been much work on phenomenology in the context of lower reheating temperatures, since all we really know is $T_{\rm reheat} \gtrsim T_{\rm BBN}$, and in such a model any particles with masses $m \gg T_{\rm reheat}$ will have Boltzmann suppressed abundances. The strongest, most robust bounds come from supernova acceleration of relic species in the galaxy \cite{Dunsky:2018mqs,Dunsky:2019api} and mean that the discovery of an FCP of mass $m$ would place an upper bound on the reheating temperature which is conservatively $T_{\rm reheat}/m \lesssim 65$ \cite{Koren:2024xof}. A lack of evidence for high-scale inflation \cite{Planck:2018jri}, as well as a lack of expected collider signatures at the electroweak scale \cite{Craig:2013cxa,Baer:2020kwz}, motivate us to revisit the notion of FCPs.

\subsection{Strategy and Takeaways}

In our previous work, we considered minimally adding to the SM a single FCP $X$ of a variety of representations. 
Since FCPs carry a quantum number that strictly no SM particles carry (electric charge mod $e$), we cannot write a symmetry invariant operator containing only one fractionally charged field. This implies that with one new such species the only possible gauge invariant interactions with SM species are the couplings to gauge bosons (and a Higgs portal coupling for a scalar). This further means that upon integrating out any FCPs and matching to the Standard Model Effective Theory (SMEFT), this quantum number must be solely internal, meaning these species only contribute to SMEFT at loop level. As a result, constraints from electroweak precision or flavor observables are automatically suppressed, meaning that colliders are necessarily the best way to probe whether such states exist in the spectrum of the Hamiltonian of the Universe.

This was borne out in our previous work on FCPs at the energy frontier, where we found that the limits depend sensitively on the quantum numbers of $X$, and can be surprisingly low---especially when $X$ has $Q/e \le 1/3$. 
We continue this pursuit here by considering models with two new FCPs, and finding that a much wider range of phenomenologies become possible. For any given SM matter field there is a family of models where it couples singly to two new such states, and in the current work we will consider a few representative examples talking to various of the SM fields. We will map out the constraints placed upon these by reinterpretations of existing searches, finding again that collider searches are most useful in line with the above general reasoning.

As in our previous work, we emphasize that the bounds we find are approximate and make various simplifying assumptions to analyze the signatures and reinterpret a variety of searches. These are necessary for us as phenomenologists in the absence of full simulation of the detector response to exotically charged matter. Our purpose is to call attention to the interesting signatures that may appear and the often-surprisingly-weak bounds which are found for these genuinely electrically charged particles. We do this in the hopes that these signatures and possible analyses garner more attention from both phenomenologists and, crucially, from experimentalists, who are, of course, the only ones who can truly understand the detector response and derive legitimately accurate bounds here.  

As stated above, particles with electric charge $Q/e \leq 1/3$ are nearly invisible in the detectors \cite{CMS:2024eyx}---but of course they do interact with matter and further experimental work on disentangling these traces from noise would be useful. Simulations of the hadronization of colored FCPs are another missing piece needed to understand their bounds more precisely, or perhaps find novel signatures in these cases. 

\begin{figure}[h!]
\centering
\includegraphics[height=7cm]{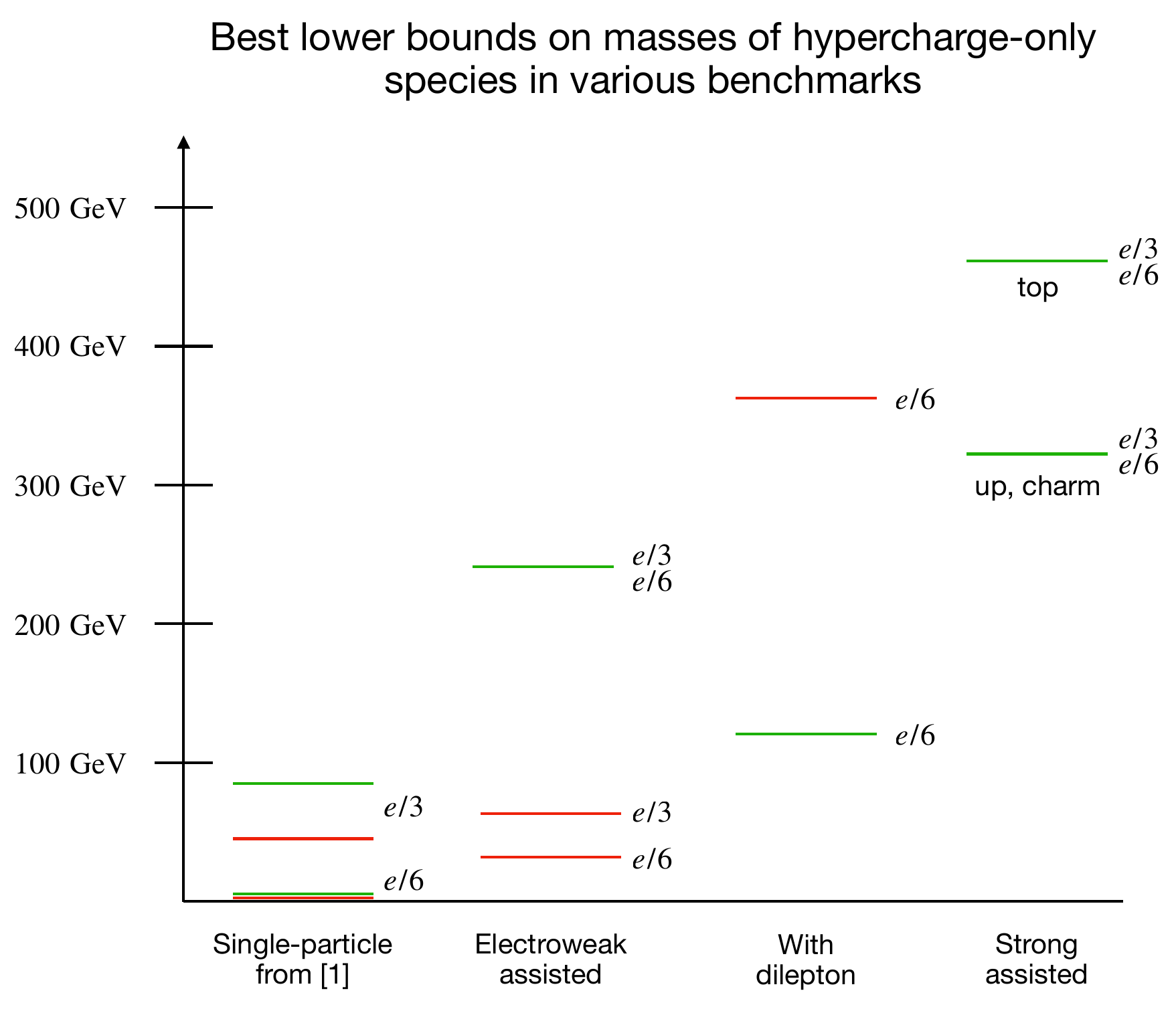} 
\includegraphics[height=7cm]{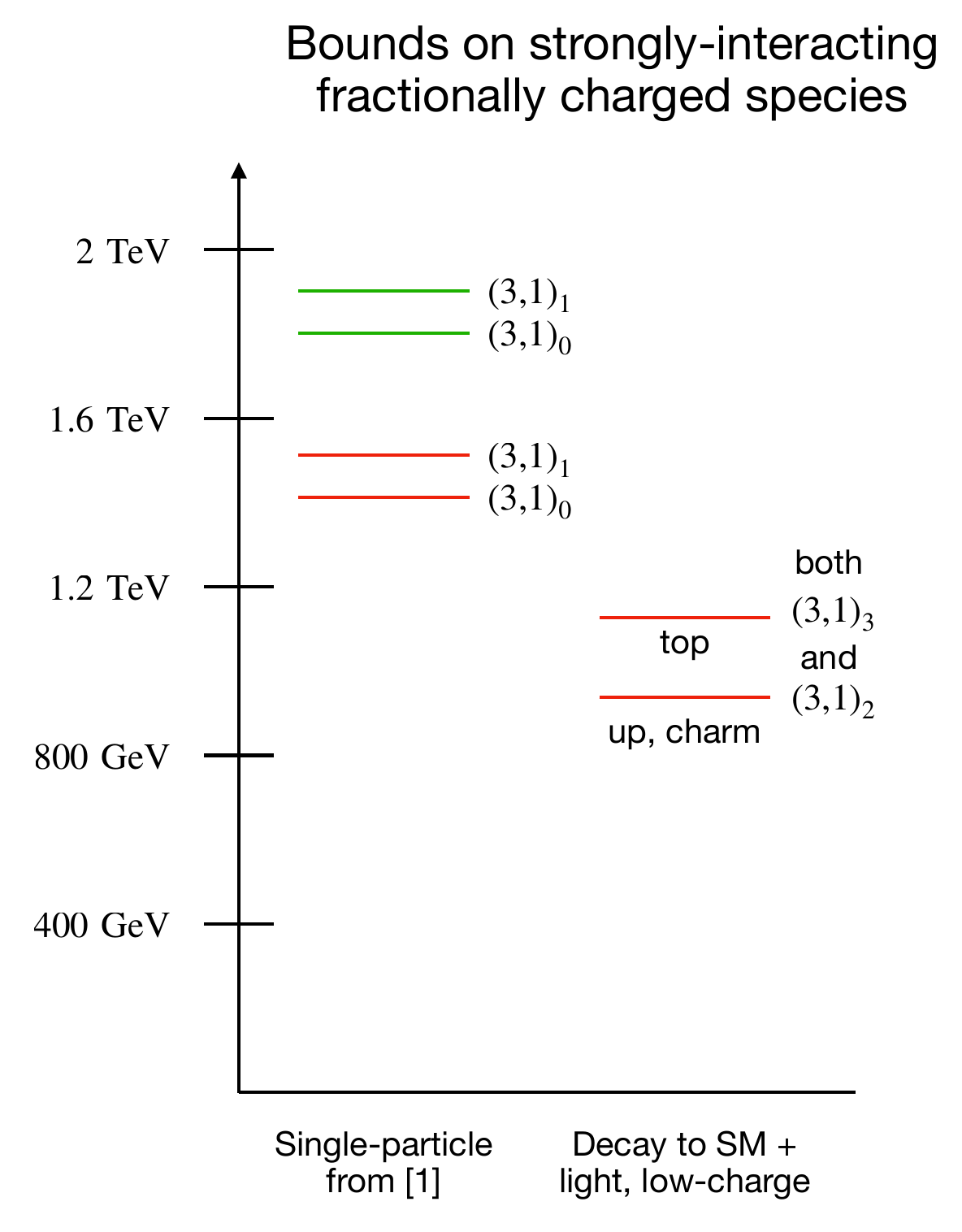} 
\caption{Comparison of constraints between single-particle extensions as studied in \cite{Koren:2024xof} and the two-particle extensions studied below, with scalars in red and fermions in green. Left: The best bounds placed upon fractionally charged species with only hypercharge, noting the improvements in two-particle extensions due to electroweak production (Secs 2.1 and 2.2) or associated production with leptons (Sec 2.3) or strong production (Sec 2.4). Model constraints which don't rely on the signature of these species are ignored. Right: The best bounds placed on strongly-interacting fractionally charged species when they are stable versus when they can decay to SM colored species and hypercharge-only fractionally charged species (Sec 2.4).}
\label{fig:boundCompare}
\end{figure}

We will find that the bounds on FCPs in two-particle extensions can be dramatically modified compared to those on single-particle extensions, as we exhibit in Figure \ref{fig:boundCompare} by comparing constraints we find below to those we found in \cite{Koren:2024xof}. We find that the stringent constraints we placed on colored FCPs can be dramatically weakened by factors of a few if they decay to SM particles + solely-hypercharged FCPs. Conversely, the extremely weak bounds placed on solely-hypercharged FCPs can become drastically stronger as their cross-sections can increase by orders of magnitude.

A general takeaway from these only-slightly-non-minimal models is that it is perfectly reasonable to expect that the lightest FCP may usually appear in events along with additional SM particles such as jets or leptons. We exhibit this explicitly in the right-hand panels of Figures \ref{fig:eRportal_flipped} and \ref{fig:uRportal} which show that the cross sections with additional leptons or jets may exceed the exclusive signature by $10^2 \sim 10^3$. This offers two particular points of guidance towards improving constraints.

One is the importance of either dedicated `fractionally charged + $\ell\bar\ell/jj$' searches, or inclusive searches for fractionally charged signatures. The CMS search \cite{CMS:2024eyx}, which often placed the leading bounds on single-particle extensions in \cite{Koren:2024xof}, selects events with only one or two energetic ($p_T > 55\, \text{GeV})$ tracks in them. In the two-particle extensions this can lead to the vast majority of signal events simply falling outside the acceptance of the search. In some cases below we will plot the bound one would find by naively reinterpreting the CMS search as if it were inclusive, but this is just to guide what sorts of constraints may be available with a more-diverse search for FCP signatures.

The second is the following `free' search strategy in the case of low charge. 
The tracks of these particles may not be reconstructed, leading to them appearing as $\slashed E_T$ in events. 
Indeed, below we have reinterpreted various such searches which include $\slashed E_T$ to place bounds on species with small fractional electric charge, assuming that there were no signatures of FCPs in the events used in the search. However, since FCPs do interact with the detector (unlike truly invisible species such as neutrinos or dark matter), these events may contain subtle signatures of FCPs. 
That is, given any already conducted search whose event criteria allowed for large $\slashed E_T$, one could now look back at these events and interrogate the tracker data for low-quality tracks and/or tracks with anomalous $dE/dx$. This could potentially find FCPs in data already on tape.

\section{Benchmark Scenarios}\label{sec:bench}

We will consider four ways in which a pair of fractionally charged states couple to a single SM particle -- four different `portals' to the SM. 
This is a subset of the possible ways two new FCPs could interact with SM states, and while we believe it largely representative of the sorts of phenomenological possibilities available, we do not here attempt to be comprehensive. 

We have discovered all of the chiral Standard Model species, as informed by the Higgs branching ratios, so new fractionally charged states must be Dirac fermions or complex scalars. We'll use $X, \bar X$ for Dirac fermions and $\phi, \phi^*$ for scalars. We will first consider Higgs couplings to a pair of either fermions or scalars, and then consider Yukawa couplings for either right-handed charged leptons or up-type quarks with a new fermion and scalar.\footnote{Of course, it is also possible to construct new Yukawa interactions for right-handed down-type quarks or neutrinos, or for left-handed quarks and leptons. We note also that one novel possibility will become available with three new scalars, namely $H \phi_1 \phi_2 \phi_3$. We leave these possibilities to future study.} We will ignore Higgs portal couplings of the form $|H|^2 |\phi|^2$, assuming that production through this coupling is subdominant to production through gauge bosons, which is true aside perhaps from extremely large Higgs portal coupling with charge $e/6$ species.

Note within each portal there are technically infinite families of possible charge assignments for $X, \phi$ that lead to fractional electric charge. We will focus on small non-Abelian representations, and on assignments where one of the FCPs is only charged under hypercharge. We do this because particles charged solely under hypercharge typically are less constrained than particles that also have $SU(2)$ and/or $SU(3)$ charges, and so offer the broadest near-term discovery potential. We'll leave hypercharge unspecified in setting up our benchmark models below, though our main interest in presenting bounds will be in assignments where one of the FCPs has $Q/e \le 1/3$. We stress again that our results will not cover all of the interesting possibilities, which are manifold.

We will use integer normalization for hypercharge, so the left-handed quark doublets $Q_i$ have hypercharge $+1$, and our convention is that the Higgs $H$ has hypercharge $+3$. For electric charges we will still normalize to the electron's charge, so that `fractionally charged' particles will have electric charges which are not an integer, $Q \in \frac{e}{6} \mathbb{Z}, \ Q \notin e \mathbb{Z}$.

\subsection{$HX_1X_2$ portal}

For our first benchmark, we consider two Dirac fermions $X_1, X_2$ coupled to the Higgs via
    \begin{align}
    \mathcal L \supset \bar X_1(i \slashed{D} - M_1)X_1 + \bar X_2(i \slashed{D} - M_2)X_2 -\lambda H \bar X_1 X_2 + \rm h.c.
    \end{align}
    We take $X_1$ to be an $SU(2)$ doublet, while $X_2$ is a singlet, 
    \begin{align}
        X_1 \sim (0,2,Y),\quad\quad X_2 \sim (0,0,Y-3)
    \end{align}
    The $X_i$ are FCPs so long as $Y \neq 3 \ (\text{mod } 6)$.
 
     The two components of $X_1$ have electric charge $(Y\pm 3)/6$; $X_2$ has electric charge $(Y-3)/6$ and will therefore mix with the lower ($t_3 = -\frac 1 2$) component of the $X_1$ doublet after EWSB (we could have coupled to $H^\dag$ rather than $H$, in which case the mixing would be between the upper component of $X_1$ and $X_2$). The spectrum therefore consists of a Dirac fermion with charge $(Y+3)/6$ and mass $M_1$, and two charge $(Y-3)/6$ Dirac fermions whose masses are skewed from $M_1, M_2$ by $\mathcal O(\lambda v)$ mixing. The mixing angle between the $(Y-3)/6$ states, 
    \begin{align}
    \tan(2\alpha) = \frac{\sqrt 2 \lambda v }{M_2 - M_1}
    \end{align}
    controls much of the collider phenomenology.

   Let us denote the $(Y+3)/6$ charged state as $X_u$ and the two $(Y-3)/6$ charged states as $X_d$ and $X'_d$, with $X_d$ dominantly composed of the $X_1$ mode and $X'_d$ the other.  The primary production modes of the $X_i$ at the LHC will be
    \begin{itemize}
    \item $\bar q q \to X_u \bar X_u, X_d \bar X_d, X'_d \bar X'_d$ via an intermediate photon or $Z$. The production of $X_d \bar X_d, X'_d \bar X'_d$ through $Z$ is $\propto \cos^2(\alpha)$ at amplitude level, and production through the photon is independent of the mixing.
    \item $\bar q q \to X_d \bar X'_d, X'_d \bar X_d$ via intermediate $Z$. This amplitude is proportional to $\sin(2\alpha)$. 
    \item $q\bar q' \to X_u \bar X_d, X_u \bar X'_d$ (or h.c). The amplitude to $X_u \bar X_d$ is $\propto \cos(\alpha)$ while the amplitude for $X_u \bar X'_d$ is $\propto \sin(\alpha)$.
    \item $gg \to X_d \bar X_d, X'_d \bar X'_d, X_d \bar X'_d, X'_d \bar X_d$ via intermediate Higgs; the first two are $\propto \sin(2\alpha)$ while the latter two are $\propto \cos(2\alpha)$ at the amplitude level.
    \end{itemize}
    
The heavier of the fractionally charged species are necessarily unstable to the lightest state, so after production will decay via emission of a $W/Z/h$. Depending on the values of $M_1, M_2$ and $\lambda\, v$, the emitted $W/Z/h$ may be on-shell.

To illustrate the effect of the portal coupling, let us consider an example with $Y=1$, giving $X_1 \sim (0,2,1)$ and $X_2 \sim (0,0,-2)$, and compare to the bounds found for both of these single-particle extensions in our previous work, Ref.~\cite{Koren:2024xof}. There, we estimated the constraints to be  $\gtrsim 1.1\, \text{TeV}$ for $X_1$ and $\gtrsim 88\, \text{GeV}$ for $X_2$. The key to the high limit on $X_1$ in~\cite{Koren:2024xof} was the long lifetime of the $Q/e = 2/3$ $X_u$ state, due to its small mass splitting ($< m_\pi$) above the $Q = -1/3$ $X_d$ component, which came only from loops of massive EW gauge bosons.

Now lets turn on the portal coupling.  Picking a mass point for concreteness, consider:
    \begin{align}
        Y = 1: M_1 = 300\, \text{GeV}, M_2 = 200\, \text{GeV},\ \lambda = 0.1
    \end{align}
The spectrum consists of a $300\, \text{GeV}, Q/e = 2/3$  fermion $X_u$, a $303\, \text{GeV}, Q/e = -1/3$ particle $X_d$, and a $197\, \text{GeV}, Q/e = -1/3$ particle $X'_d$. Compared to the scenario without a portal coupling, the setup with $\lambda \bar X_1 X_2$ has more production channels, such as $gg \to \bar X_i X_i$ through a Higgs. However, the portal coupling provides additional splitting, which at this example point is now $106\, \text{GeV} \gg m_\pi$, so that $X_u \to X_d'$ will be prompt. For this scenario, all $X_u$ and mixed $X_d - X_u$ production modes lead to pairs of $X'_d$ along with $\ell^\pm \nu$, $\bar q q$ (or pairs thereof) from $X_u$ beta decay. Similarly, all $X_d$ decay promptly to $X'_d$ by emitting an (off-shell, for the numbers above) Higgs or $Z$, leading to $X'_d + \ell^+\ell$, $X'_d + \bar q q$, etc. As $X'_d$ tracks are effectively invisible in the detector, constraints on this scenario are set by searches for leptons plus $\slashed E_T$ or jets plus $\slashed E_T$, which leads to far worse constraints on $X_1$. 

The hierarchy $M_1 < M_2$ can lead to even sneakier signals. Swapping $M_1 \leftrightarrow M_2$ in the numerical example, $X_u$ has mass $200\, \text{GeV}$ rather than $300\, \text{GeV}$, and the masses of $X_d$ and $X_d'$ are swapped. A mass splitting of $3\,\text{GeV}$ between $X_u, X_d$ is still big enough that $X_u$ will decay promptly, but now its beta decay products will be much softer and therefore more difficult to detect. If we treat the beta decay products as invisible (to first approximation), then producing either $X_u$ or $X_d$ leads to no visible signatures in the tracker.

In addition to these direct production signatures, the coupling between FCPs and the Higgs introduces further low-energy phenomenology past that in the single-particle extensions. Firstly, the only electroweak precision observables (EWPO) generated by gauge couplings alone are $W$ and $Y$, but once we include the Higgs coupling there will be a contribution to the $\hat S$ and $\hat T$ parameters as well. Following Ref.~\cite{Barbieri:2004qk,Wells:2015uba} to convert between dimension-six operators and the conventional EWPO, we find that, in the limit of degenerate $X$ masses:
\begin{align}
    \hat S \sim \frac{\lambda^2\, g^2\, v^2\, (11-20\,Y)}{960\, \pi^2\, M^2_X}, \quad\quad \hat T \sim \frac{\lambda^4\, v^2}{40\, \pi^2\, M^2_X}
\end{align}
For unequal masses, the expressions are more complicated but can be extracted using the matching program {\tt Matchete}~\cite{Fuentes-Martin:2022jrf}. These do not provide strong bounds for the values of $Y$ we are interested in unless $\lambda$ is large; for example, the PDG fit has $\hat{T} = (0.00 \pm 0.06)\alpha$ \cite{ParticleDataGroup:2024cfk} (having set $U=0$), which translates into $\lambda^4 \frac{v^2}{M_X^2} \lesssim 1.5$---a weak bound as generally expected above since FCPs contribute to SMEFT only at loop level. 

Second, if one or more of $X_d, X_d'$ is lighter than half of the mass of the Higgs, the Higgs can decay to it. Assuming that only one state is kinematically open for two-body Higgs decay, the partial width $h \to \bar X_i X_i$ is
\begin{align}
    \Gamma(h \to \bar X_i X_i) = \frac{\lambda^2 \sin^2(2\alpha)\, m_h}{16\pi}\Big(1 -\frac{4M^2_X}{m^2_h }\Big)^{3/2}
\end{align}
where $M_X$ is the mass of $X_i = \{X_d, X'_d$\} (if both $X_d, X'_d$ are light enough, there are additional terms $h \to \bar X_i X_j$ $\propto \cos^2(2\alpha)$). Comparing the $h \to \bar X_i X_j$ partial width with the current Higgs invisible width ($\Gamma_{h \to iv.} \lesssim 0.11\, \Gamma_h$ \cite{ParticleDataGroup:2024cfk}), we can bound the light-$X$ corner of parameter space.

The total cross section for $pp \to X_{d,\ell}\bar X_{d,\ell} + \cdots$ is shown below in Figure~\ref{fig:HPportal} as a function of $M_1$ and $M_2$ for a couple benchmark choices of Y and $\lambda$. Here, $X_{d,\ell}$ indicates the lighter of $X_d, X'_d$ (which depend on the ordering of $M_1$ vs. $M_2$) and the $\cdots$ are any combination of $W/Z/h$ decay products. The cross sections are in $pb$ and were calculated for a $13\, \text{TeV}$ center of mass LHC. All contributions were calculated to lowest order only, using {\text NNPDF30nlo} parton distribution functions~\cite{Ball_2015,Hartland_2013} with $\alpha_s = 0.118$ and a factorization scale $\mu^2_F = \hat s$.

    \begin{figure}[t!]    
    \centering
    \includegraphics[width=0.45\textwidth]
    {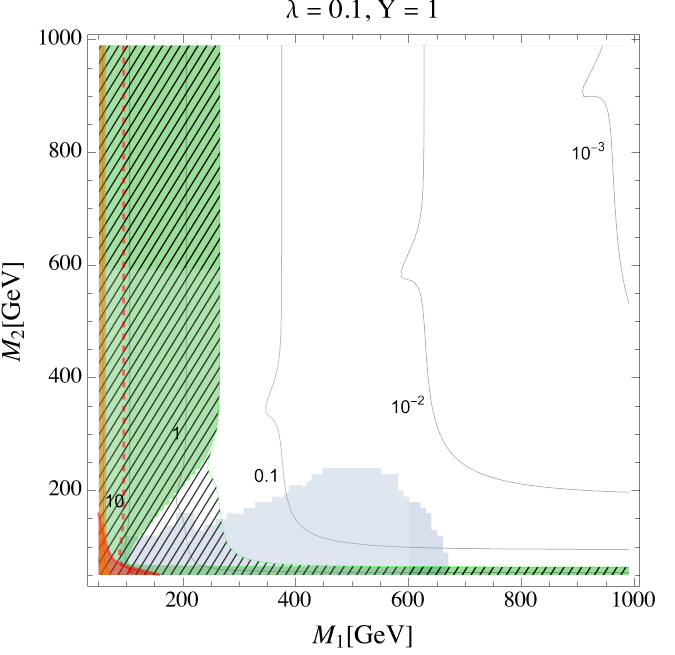}
      \includegraphics[width=0.45\textwidth]
         {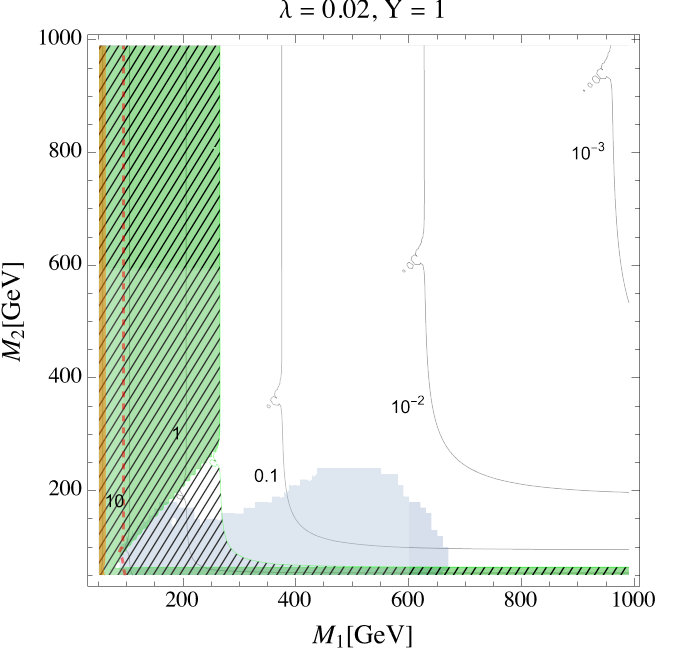}\\
         \vspace{0.4cm}
         \includegraphics[width=0.45\textwidth]
         {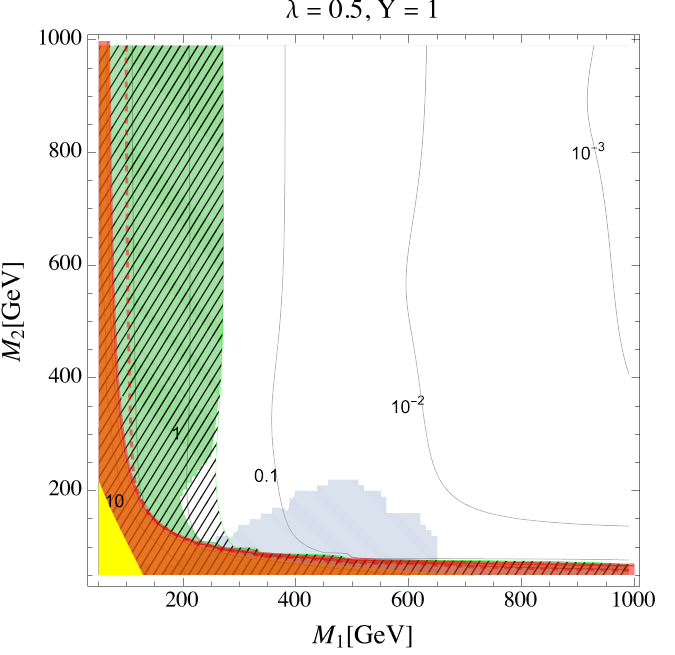}
          \includegraphics[width=0.45\textwidth]
         {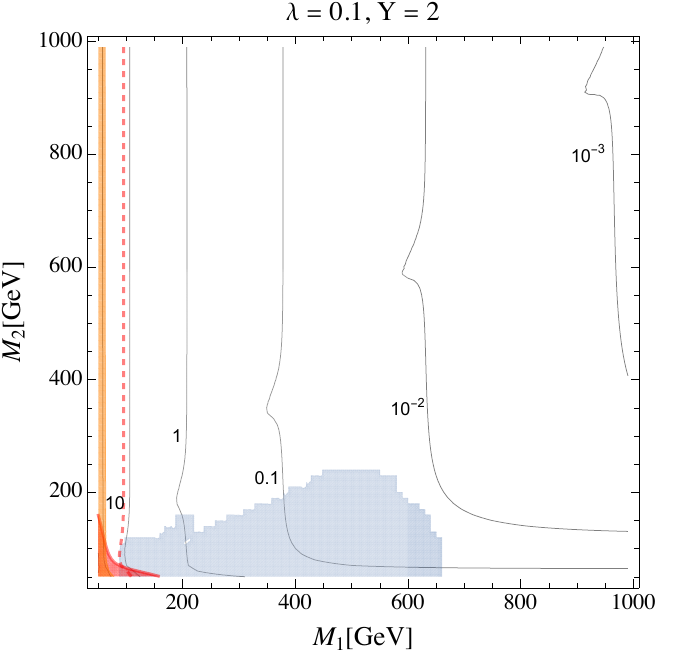}
    \caption{Total cross section $pp \to \bar X_i X_j, X_i = X_u, X_d, X'_d$ at the LHC. The top two panels and the bottom left panel show $Y=1$ -- so that $Q_{X_d} = Q_{X'_d} = 1/3$ (matching one of the scenarios in Ref.~\cite{Koren:2024xof}) --  for various $\lambda$, while the bottom right panel shows $Y=2$ ($Q_{X_d} = Q_{X'_d} = -1/6$) for $\lambda = 0.1$. The green band corresponds to the limits from the CMS search for $|Q/e|=1/3$ particles, Ref.~\cite{CMS:2024eyx}. The hatched region shows an estimation CMS limit if the restriction to one or two tracks is removed, see text for more details. There is no limit from~\cite{CMS:2024eyx} when $|Q/e|= 1/6$. The gray shaded area corresponds to the bounds from $W+Z+\slashed E_T$ searches, where we have treated the lightest $X_d/X'_d$ state as if it were invisible, and the orange (dashed red) are the current (projected) milliQan bounds. The red band in the lower left corner is the bound from invisible Higgs decay. As the partial width $\sim \lambda^2$, this bound is most visible when $\lambda = 0.5$. The $\lambda = 0.5$ scenario is also the only place where EWPO constraints ($2\sigma$ $T$ parameter exclusion), shown in the yellow, appear.}
    \label{fig:HPportal}
    \end{figure}
 
We show $Y=1$, so that the quantum numbers of $X_1$ correspond to a multiplet studied in~\cite{Koren:2024xof}, and $Y=2$ -- so that the lightest charge states in the setup will have $|Q/e| = 1/6$. We see that the contours of the cross section are nearly vertical, except when $M_2$ is small. This shape occurs because $pp \to \bar X_i X_j$ production for $|Q_X/e| < 1$ is dominated by the states which have $SU(2)$ charge~\cite{Koren:2024xof} -- the doublet $X_u, X_d$, which has mass $\sim M_1$.\footnote{The production of particles charged only under hypercharge is always $\sim Q^2_X$ and therefore suppressed when $Q_X$ is small, while the production of particles charged under $SU(2)$ has terms that are independent of $Q_X$.} The contours shift from vertical to more horizontal at low $M_2$, where the lightness of the singlet particle in the scenario offsets its suppressed production.

Bounds from collider searches, the Higgs partial width, and EWPO have been superimposed on top of the cross-section contours. When $M_1 > M_2$, such that the predominantly doublet states $X_u, X_d$ are heavier than the predominantly singlet state, there are bounds from searches in $W^\pm + Z + \slashed E_T$ final states~\cite{CMS:2024gyw,ATLAS:2024qxh}. 
These searches are designed to look for the charged current production of a chargino and heavy neutralino that subsequently decay to $W/Z/h$ plus the lightest neutralino. However, if we assume $X_{d,\ell}$ is effectively invisible in the detector, the same search can be applied to the Higgs portal scenario. The search is most efficient when the doublet states and singlet are well separated ($M_1 \gg M_2)$, so that the $W/Z/h$ decay products are energetic, but there are variations of the electroweakino search that look for more degenerate spectra that we port to our scenario. When $M_2 > M_1$ the predominantly singlet state is the heaviest. It will decay to $X_u$ and $X_d$ plus a $W/Z/h$, so Drell-Yan production $pp \to X'_d \bar X'_d$ can lead to the same final state searched for in~\cite{CMS:2024gyw,ATLAS:2024qxh}. However, both the production rate and the branching ratio are significantly lower than when $M_1 > M_2$. 

The bounds from the electroweakino searches are shown in the blue regions in Fig.~\ref{fig:HPportal}. To extract the bounds, we compare the $95\%\, \text{CL}$ cross section from~\cite{CMS:2024gyw} (via HEPdata) with the $pp \to W+Z + X_{d,\ell} \bar X_{d, \ell}$ cross section. As expected, the excluded region is where $M_1 > M_2$ and extends to roughly $650\, \text{GeV}$ for light $X_d$. Comparing the panels, we see that this bound is not strongly affected by $\lambda$ or $Y$. References~\cite{CMS:2024gyw,ATLAS:2024qxh} also place constraints on the $W+h + \slashed{E}_T$ final states, but we find that these bounds are weaker for our set-up. 

Another relevant collider bound comes from the CMS search for FCPs, Ref~\cite{CMS:2024eyx}. This search was reviewed in our previous work, and sets bounds for $|Q/e| \ge 1/3$. As mentioned earlier, the CMS search selects events with only one or two energetic tracks. This means that events such as $pp \to \bar{X}_u X_d \to W \bar X_d X_d$ (where we take $X_d$ to be the lightest particle) which contain three or four charged tracks (two of which have low $dE/dx$) may not be retained. 

If the additional tracks are soft, i.e. when the mass difference between $X_u$ and $X_d$ in the example above is small, the CMS search will retain these events and be able to constrain such a scenario. The larger the mass difference becomes, the more likely it is for events to be rejected by the selection cut on extra energetic tracks. However, a further confounding factor is that the track reconstruction efficiency for $Q/e = 1/3$ is itself quite low, and so for this benchmark events would largely need to have \textit{two} extra energetic tracks in order to be cut.

To account for the approximate impact of extra tracks, we present the CMS bound two different ways. As a conservative bound (green shaded region in Fig.~\ref{fig:HPportal}), we consider the production of the lightest multiplet. For $M_1 > M_2$, this means we only consider $X'_d$ production, while for $M_2 > M_1$ we consider production of both $X_u, X_d$ (the $X_1$ multiplet), given that $X_u - X_d$ mass difference is small (at least for small $\lambda$). In either regime, we compare the production of the lightest multiplet with the $|Q/e|= 1/3$ cross-section limit from~\cite{CMS:2024eyx}. As a more optimistic bound (hatched region in Fig.~\ref{fig:HPportal}), we apply the CMS bound to the inclusive $pp \to \bar X_i X_j$ cross section, where signal events with three or more tracks are included. The two variations on the bound overlap for most of parameter space, but there is some difference for $M_1 \sim M_2$. 
We expect that the true CMS bound lies somewhere between the green and hatched contours, but precise reinterpretation relies on fuller simulations of detector response.  

The last collider bound we include is the limit and projection from the milliQan experiment~\cite{milliQan:2021lne}. We assume a slightly different setup than milliQan -- they assume FCPs couple only to the photon -- so we need to first reverse engineer the cross section from the mass vs. charge curves in~\cite{milliQan:2021lne}. This reverse engineering yields a current cross section bound of $68\, \text{pb}$ and a projected bound of $14.5\, \text{pb}$.
As with the rest of our bounds, this method is an approximation as it assumes the acceptance is independent of the mass.\footnote{We find that the mass dependence of the acceptance, at least from a purely kinematic standpoint, is negligible.} We also assume that additional objects in the event do not affect milliQan's signal efficiency, so we can apply the bound to the total $pp \to \bar X_i X_j$ cross section.

With these assumptions, the projected milliQan bounds reach roughly $M_1 = 100\, \text{GeV}$. This mass is larger than the mass limit projected in~\cite{milliQan:2021lne} because, as mentioned earlier, we assume FCPs with both $SU(2)$ and $U(1)$ charges, which raises their production cross-section. For $Y=1\, (Q_X = 1/3)$, the milliQan bounds and projection are superseded by the CMS FCP search. Note that, for $Y=2\, (Q_X = 1/6)$, there is no LHC bound. 

Summarizing, we see that the bounds on a particle with quantum numbers $\sim (0,2,1)$ (mass $M_1$ in our setup) can be significantly weakened when we include the portal interaction, dropping from $\sim 1.1\, \text{TeV}$ to $200 - 700\, \text{GeV}$ depending on the mass of the $SU(2)$ singlet it couples to. Weaker bounds on $(0,2,1)$ come at the expense of tighter constraints on its singlet partner, $(0,0,-2)$. In addition to drastically weakening the bounds on the electroweak-charged $(0,2,1)$, we find also that the bound on the singlet $(0,0,-2)$ can be strengthened enormously, from $\sim 90\, \text{GeV}$ to $\sim 250\, \text{GeV}$. If $Y=2$, such that $|Q/e|$ of the singlet is $1/6$, the current bounds are even looser. In both charge assignments, there are sizable regions of open parameter space with large-ish, $\mathcal O(0.1-\text{few\, pb})$ cross section. 

\subsection{$H\,\phi_2\, \phi^2_1$ portal}\label{sec:hphi1phi2}

For our second benchmark, we consider a variation of the Higgs portal involving two scalar FCPs. We can keep the coupling marginal -- thus avoiding the introduction of another mass scale -- if one of the fractionally charged scalars appears twice in the interaction. Specifically, we will consider 
\begin{align}
\mathcal L \supset \lambda H\, \phi_2\, \phi^2_1 + h.c.
\end{align}
where $\phi_1, \phi_2$ are scalars and $\phi_2 \sim  (0,2,-(3+2Y))$, $\phi_1 \sim (0,0,Y)$.  Note that the singlet is fractionally charged only if $Y \neq 0 \ (\text{mod } 6)$ and the doublet is fractionally charged only if also $Y \neq 3 \ (\text{mod } 6)$, so we impose both of these. We will refer to the two components of $\phi_2$ as $\phi_{2,u}$ ($t_3 = 1/2)$ and $\phi_{2,d}$ ($t_3 = -1/2)$.

As the interaction involves four fields, there is no mass mixing among the FCPs in the EW broken phase. As a result, there are fewer production channels to consider compared to the previous benchmark. At the LHC, $\phi_1$ and both components of $\phi_2$ 
multiplets are produced via Drell-Yan, while production via $W^{\pm}$ exchange leads to $\phi^*_{2, u} \phi_{2,d} + h.c.$. There will also be gluon fusion through an $s$-channel Higgs directly into $\phi_{2,u}\, \phi^2_1$ that is proportional to the portal coupling $\lambda$.

Let us begin with a setup with $Y=1$, so that $Q_{\phi_1}/e = 1/6$, $Q_{\phi_{2,u}} = -1/3$ and $Q_{\phi_{2,d}} = -4/3$. The collider signature for this benchmark depends on $M_{\phi_2}/(2 M_{\phi_1})$. If $M_{\phi_2} > 2 M_{\phi_1}$, all $\phi_2$ will decay into $\phi^*_1$; $\phi_{2,u} \to (\phi^*_1)^2$ directly through the portal coupling, while $\phi_{2,d}$ must first beta decay to $\phi_{2,u}$, so $\phi_{2,d} \to W^{(*)}(\phi^*_1)^2$, where $W^{(*)}$ represents the decay products of an (off-shell) $W^\pm$. Following the steps laid out in Ref.~\cite{Koren:2024xof} to determine the mass splitting and lifetime, we find that for $Y=1$, $M_{\phi_2,d} - M_{\phi_2,u} \sim 0.3 - 0.6\, \text{GeV}$ (as this is generated solely by loops of $W, Z$), which translates to a sub-cm decay length. Therefore, for this assumption of charges and masses, the production modes and final states are:
\begin{itemize}
\item Via Drell Yan: $pp \to \phi_1 \phi^*_1$, $pp \to \phi^*_{2,u} \phi_{2,u} \to 2 \phi_1 2 \phi^*_1$, $pp \to \phi^*_{2,d} \phi_{2,d} \to 2W^{(*)} + 2 \phi_1 2 \phi^*_1$.
\item Via charged current: $pp \to \phi^*_{2,u} \phi_{2,d} + h.c. \to W^{(*)} + 2 \phi_1 2 \phi^*_1$.
\item Via gluon fusion $pp \to \phi_{2,u} \phi^2_1 \to 2 \phi_1 2\phi^*_1$. If $M_{\phi_1} < m_H/4$, this process is a rare Higgs decay.
 \end{itemize}
For this choice of $Y$, $\phi_1$ is essentially invisible at the LHC. The companion states $\phi_{2,u}, \phi_{2,d}$ have larger charge, but decay promptly. 

The decay of $\phi_{2,d}$ into $\phi_{2,u}$ produces hadrons or leptons plus a neutrino, though these will be soft given that $M_{\phi_2,d} - M_{\phi_2,u} \lesssim \text{GeV}$.  The net signal is two or more $Q/e = 1/6$ particles plus soft pions/lepton+neutrino pairs from $W^*$ decays. In this case, the strongest bound comes from the production of the $|Q/e| = 4/3$ state $\phi_{2,d}$. $\phi_{2,d}$ beta decays to $\phi_{2,u}$, however, for sufficiently small $M_{\phi_2}$, $\phi_{2,d}$ may live long enough to appear in disappearing track searches. Comparing the $\phi_{2,d}$ production cross-section (as a function of $M_2$ and the $\phi_{2,d} - \phi_{2,u}$ mass difference) to the limits shown in Ref.~\cite{CMS:2023mny}, we find a small region at low $M_{\phi_2}$ is excluded. The present bound is weaker than in~\cite{CMS:2023mny} because the mass splitting is larger.

If $M_{\phi_2} < 2 M_{\phi_1}$, $\phi_{2,u}$ is stable.\footnote{The fact that $\phi_2$ can be exactly stable even in a region where it is not the lightest fractionally charged particle ($M_{\phi_1} < M_{\phi_2} < 2 M_{\phi_1}$) may give pause, but can be understood as follows. From the charge assignment of $\phi_2$ we see it is charged under the $\mathbb{Z}_3$ center but not the $\mathbb{Z}_2$ center, whereas for odd $Y$ we have that $\phi_1$ is charged under both $\mathbb{Z}_{2}$ and $\mathbb{Z}_3$. Each factor imposes its own selection rule, so both can be stabilized by their fractional charge in the region where $\phi_2$ is the lightest state charged under only $\mathbb{Z}_3$ and $\phi_1$ is the lightest state charged under $\mathbb{Z}_2$.} All $\phi_2$ production modes lead to pairs of $|Q/e| = 1/3$ particles, which can be bounded by the CMS FCP search in Ref.~\cite{CMS:2024eyx}.

The total cross section for all $\phi_i$ particles is shown below in Fig.~\ref{fig:hphi2phi12} as a function of $M_{\phi_1}$ and $M_{\phi_2}$. We have assumed that the portal coupling $\lambda$ is small enough that $pp \to h \to \phi_2 \phi^2_1$ is negligible. We also overlay the bounds from the CMS fractional charged particle search for $|Q/e| = 1/3$ along with the current and projected milliQan bounds. Note that the CMS bound only applies when $M_{\phi_2} < 2 M_{\phi_1}$, where $\phi_{2,u}$ is stable. It is derived by comparing the production cross section of at least one $\phi_{2,u}$ state ($\sigma(pp \to \phi^*_{2,u}\phi_{2,u}) + \sigma(pp \to \phi_{2,u}\phi^*_{2,d})+ c.c.)$ to the CMS limit $0.39\, \text{pb}$~\cite{CMS:2024eyx}. We include production of $\phi_{2,d}$ when implementing the CMS bound, even though $\phi_{2,d}$ events will potentially have additional tracks from the $\phi_{2,d}$ beta decay to $\phi_{2,u}$. These additional tracks will be soft, given the small mass difference between the two $\phi_2$ states, so, as in the previous setup, we assume they will not interfere greatly with the CMS signal selection. 

\begin{figure}[h!]
\centering
\includegraphics[width=0.48\textwidth]{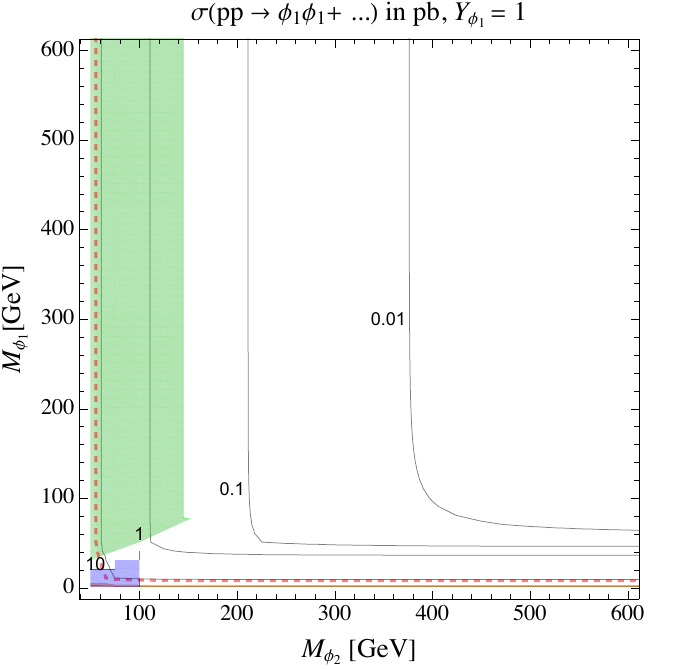} 
\includegraphics[width=0.48\textwidth]{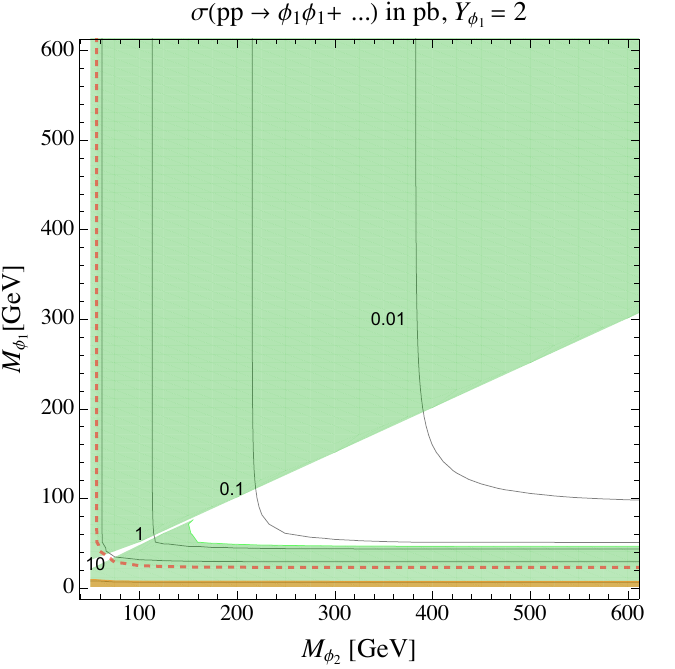} 
\caption{The LHC cross section (in pb) of $pp \to \phi_1\phi^*_1 + X$ where $X$ is any combination of $W^{(*)}$ decay products and additional $\phi_1 \phi^*_1$. The left panel shows the scenario where $Y=1$, so that $Q_{\phi_1}/e = 1/6$, while the right panel shows $Y = 2$, $Q_{\phi_1}/e = 1/3$.
Here we have assumed that $\lambda$ is small enough that $pp \to h \to \phi_2 \phi^2_1$ is negligible. The shaded green is excluded by the CMS FCP search~\cite{CMS:2024eyx}. In the left panel, this bound is only applicable when $\phi_{2u}$ is stable, while in the right panel it can be applied everywhere. The reinterpreted disappearing track bounds from Ref.~\cite{CMS:2023mny} are shown in blue (lower left corner only), and the orange and red dashed are current/projected milliQan bounds. }
\label{fig:hphi2phi12}
\end{figure}

For low $M_{\phi_1}, M_{\phi_2}$, this scenario can also be constrained by the exotic Higgs decay $h \to \phi_2 \phi^2_1$. Comparing the rate for this three-body decay to the limit on the Higgs invisible width, we find that this can bound scenarios where $\lambda$ is large. However, the bounds are always weaker than the bounds from other collider probes.

Finally, let us repeat the analysis for $Y=2$. The production modes and final states remain the same as in the $Y=1$ scenario, but the electric charges of the $\phi$ are different. If $M_{\phi_2} > 2 M_{\phi_1}$, all production modes now lead to pairs of $|Q/e| = 1/3$ $\phi_1$ particles. Similarly, if $M_{\phi_2} < 2 M_{\phi_1}$ the stable $\phi_{2,u}$ has charge $Q/e = -2/3$. Both regimes are subject to the bounds from~\cite{CMS:2024eyx}. The bounds are shown in the right panel of Fig.~\ref{fig:hphi2phi12}, overlaid on top of the inclusive LHC $pp \to \phi_1 \phi^*_1 + X$ cross-section. Note that the cross-section contours are nearly identical to those in the left panel, but the constraints from Ref.~\cite{CMS:2024eyx} are now significantly stronger, ruling out the $M_{\phi_2} < 2 M_{\phi_1}$ parameter space for the range of masses shown.

We see that, similarly to the first benchmark, adding the portal interaction can enormously strengthen the bound on a singlet FCP while greatly loosening the bound on the electroweakly charged particle. Specifically, in Ref.~\cite{Koren:2024xof}, we found no LHC bound on scalars with $|Q_X/e| = 1/6$. Once coupled to a heavier, larger charge state, we can bound the setup (weakly) using disappearing tracks up to $\sim 30\, \text{GeV}$. For this same region of parameter space ($M_2 > 2\, M_1$), the $SU(2)$ doublet $\phi_2$ can decay and the bounds from~\cite{CMS:2024eyx} no longer apply. The impact of the $\phi_2 \to (\phi^*_1)^2$ decay is even more striking if $Q_{\phi_{2,u}}/e = 2/3$ (right hand panel of Fig.~\ref{fig:hphi2phi12}), where the CMS bound would be strongest if not for the decay.   

\subsection{$e_R$ portal}

In the next two benchmarks we consider models where two FCPs couple to a SM fermion. In order for a Lorentz invariant operator to exist, one of the new particles must be a fermion and the other a boson, which we take to be a scalar. We denote the (complex) scalar as $\phi$, and use $X$ for the Dirac fermion.

For our first example, we take the portal fermion -- the fermion involved in the coupling to FCPs -- to be a right-handed lepton, so that the Lagrangian contains
 \begin{align}
 \label{eq:erportal}
 \mathcal L \supset \bar X(i \slashed{D} - M_X)X + |D_\mu \phi|^2 - M^2_\phi |\phi|^2 - y \bar e P_L X\, \phi + \rm h.c.
 \end{align}
Here we take $\phi$ to have charges $(0,0,Y)$ while $X \sim (0,0,6-Y)$, with $Y \neq 0 \ (\text{mod } 6)$ so that the electric charges of $\phi$ and $X$ are fractional. For simplicity, we consider coupling to only one of the charged lepton mass eigenstates.

Both $\phi, X$ will be produced via Drell-Yan, but, because of the portal term, the heavier of $\phi, X$ will decay to a lepton plus the lighter state. As long as $\phi$ and $X$ are not nearly degenerate and $y$ is not infinitesimal (as an example: $y \gtrsim 10^{-7}\, \sqrt{\text{TeV}/M_\phi}$ for heavy $\phi$ decaying to massless $X$), the decay will be prompt.

\begin{figure}[h!]
\centering
\includegraphics[width=0.49\textwidth]{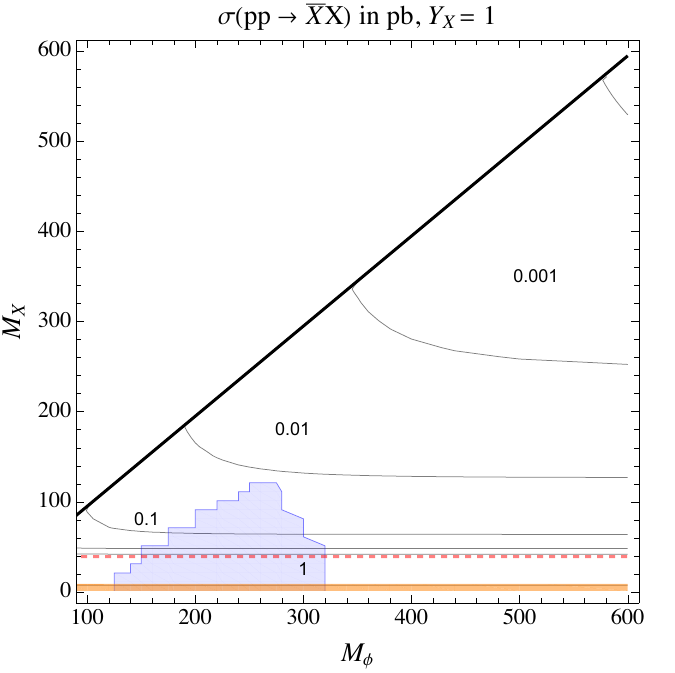}
\includegraphics[width=0.49\textwidth]{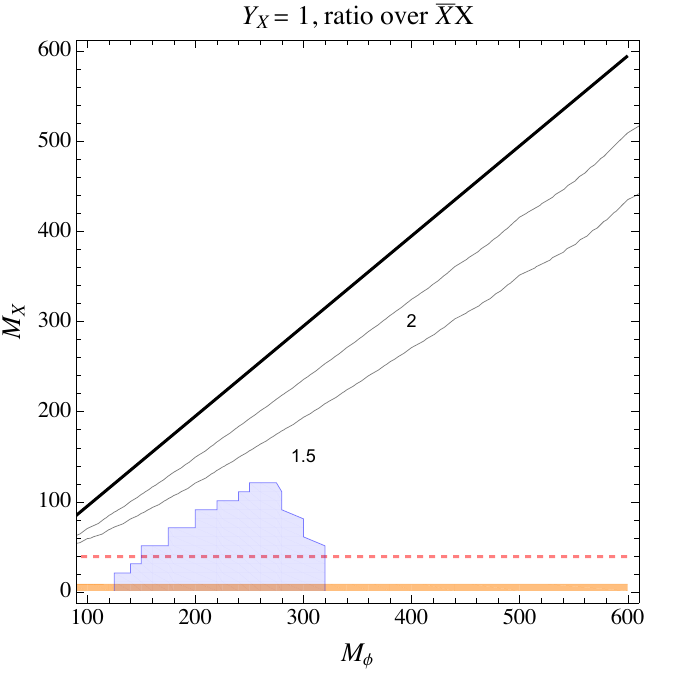}
\caption{Total cross-section for $pp \to \bar X X$ + $pp \to \bar X X + \ell^+\ell^-$ as a function of the FCP masses $M_\phi$, $M_X$. We assume $M_\phi > M_X$. The shaded portion in the lower left is the excluded region from $\slashed E_T +$ dilepton searches at the LHC for a single flavor right handed slepton, assumed to decay to lepton plus neutralino and with all other superparticles decoupled. In the right panel, the ratio of the cross section to $pp \to \bar X X$ and $pp \to \bar X X + \ell^+\ell^-$ to the cross section to $pp \to \bar X X$ alone. }
\label{fig:eRportal}
\end{figure}

Let us first consider the case where $M_\phi > M_X$ and $Y = 5$, so that $X$ has an electric charge of $Q/e = 1/6$. Assuming prompt decays, the collider signals will be $pp \to \bar XX$ plus $pp \to \ell^+\ell^- + \bar X X$ where, at least in the minimal setup above, the leptons have the same flavor. The total cross section $\sigma(pp \to \bar XX) + \sigma(p p \to \bar X X + \ell^+\ell^-)$ at the LHC is shown in Fig.~\ref{fig:eRportal}. Along with the total cross section, we also show the ratio of total inclusive (meaning plus additional leptons) $\bar XX$ cross-section over pure Drell-Yan $pp \to \bar X X$ alone. This ratio shows the `boost' in the overall $X$ production rate in this scenario, ignoring any other objects in the event.

On top of these cross-section and ratio contours, we have overlaid the bounds from~\cite{CMS:2024gyw}, which searched for sleptons decaying to neutralinos plus leptons in the same-flavor, opposite charge dilepton plus $\slashed E_T$ final state.\footnote{ATLAS bounds on this scenario are similar, see Ref.~\cite{ATLAS:2019lff}} In applying this bound, we are treating the charge $Q/e = 1/6$ $X$ particles as effectively invisible. To find the contours, we extracted the cross section limits on purely right-handed selectrons from Ref.~\cite{CMS:2024gyw} HEPdata as a function of the $X$ mass, then solved for the $\phi$ mass that corresponded to the limiting cross-section. We need to solve for the $\phi$ mass because the $\phi$ has electric charge $Q/e = 5/6$, while sleptons have $Q/e = 1$. From previous searches which differentiated slepton flavors~\cite{ATLAS:2019lff, ATLAS:2019gti}, we expect the bounds for portal couplings involving electrons or muons to be similar, with bounds on portals to $\tau_R$ somewhat weaker. We have also overlaid the current and anticipated cross-section limits on a $Q/e = 1/6$ particle from milliQan.

Examining the right panel of Fig.~\ref{fig:eRportal}, we see that the boost to the $X$ cross section provided by the additional $\phi$ production mode is not significant. This may seem strange at first, especially for those familiar with supersymmetry, where slepton production can boost the production of neutralinos. The reason why the enhancement is smaller in our case compared to supersymmetry is that $X$ is electrically changed (while the neutralino obviously is not). Being charged, $X$ can be produced from $\bar q q \to \gamma^* \to X\bar X$. When $X$ is light, that photon-induced cross section is large, and the addition of $pp \to \phi^*\phi \to \bar X X + \ell^+\ell^-$ is a small effect. This logic also explains why the cross section contours in the left panel are mainly horizontal, despite the fact that $Q_X$ is smaller than $Q_\phi$.
 
Let us now consider the opposite mass hierarchy, where $M_X > M_\phi$ and $Y=1$, so that $\phi$ is the lighter, $Q= 1/6$ state. The experimental signature is the same -- a pair of `invisible' $Q/e = 1/6$ particles, either alone or produced along with a pair of same-flavor, opposite sign leptons. The total cross section $\sigma(pp \to \phi^*\phi) + \sigma(pp \to \phi^*\phi + \ell^+\ell^-)$ is shown below in Fig.~\ref{fig:eRportal_flipped} as a function of $M_\phi$ and $M_X$, along with the ratio of the inclusive $\phi^*\phi + \cdots$ cross section to $pp \to \phi^*\phi$ alone.  

\begin{figure}[h!]
\centering
\includegraphics[width=0.49\textwidth]{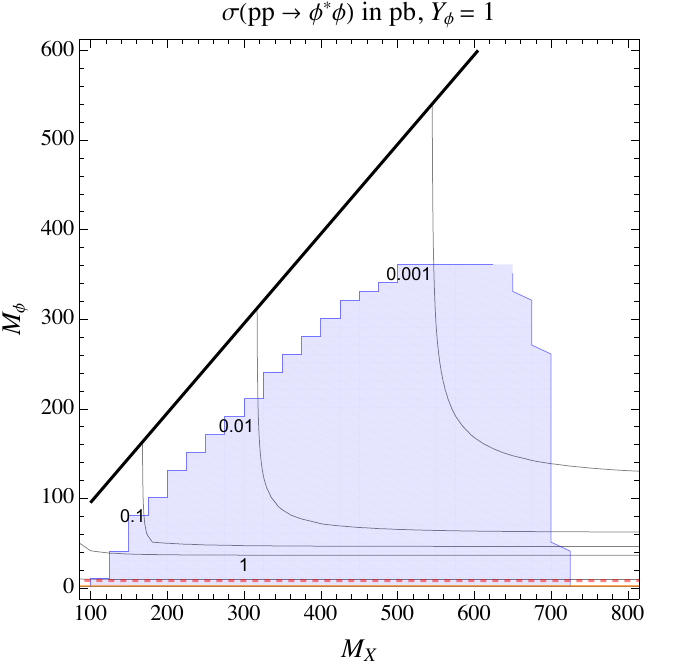}
\includegraphics[width=0.49\textwidth]{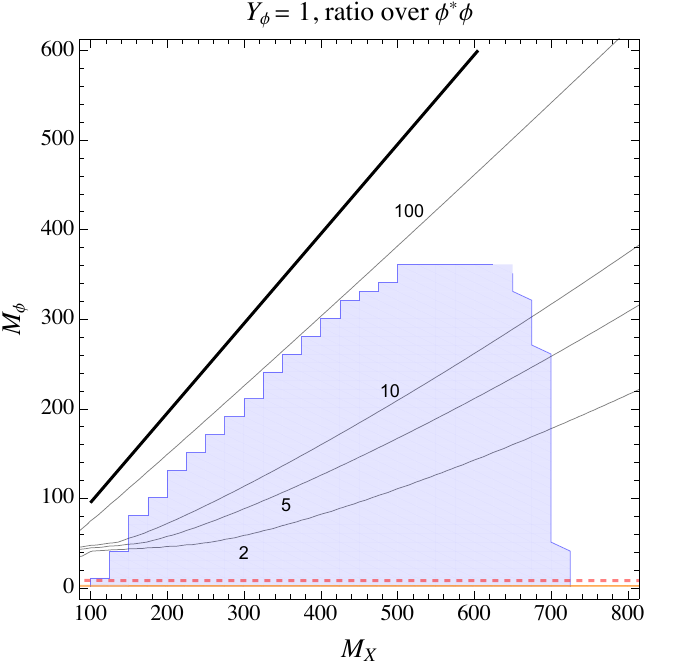}
\caption{As above, except now $M_\phi < M_X$ and we pick $Y$ such that $\phi$ has electric charge $1/6$. Note that $(\sigma(pp \to \phi^* \phi) + \sigma(pp \to \phi^*\phi + \ell^+\ell^-))/\sigma(pp \to \phi^*\phi)$ is significantly larger than in the case where the charge $1/6$ state is a fermion. }
\label{fig:eRportal_flipped}
\end{figure}

While the experimental signature is the same, the statistics of the particles have been flipped so the heavy, large charge particle is a Dirac fermion while the light particle is a scalar. Assuming equal charges and masses, the cross section for a pair of Dirac fermions is roughly an order of magnitude larger than for a pair of charged scalars, because fermions have more degrees of freedom and their interaction with gauge bosons is momentum-independent. This difference is reflected in Fig.~\ref{fig:eRportal_flipped}, as the cross section contours are nearly vertical once $M_\phi \gtrsim 100\, \text{GeV}$. The effect is especially pronounced in the right hand panel, as the numerator ($pp \to \bar X X$ vs $pp \to \phi^*\phi$ in the case above) is larger, while the denominator (now $pp \to \phi^* \phi$ vs. $pp \to \bar X X$ previously) is smaller.

As before, we have overlaid the bounds from $\ell^+\ell^- + \slashed E_T$ (slepton) searches from Ref.~\cite{ATLAS:2019lff} and from the milliQan experiment. The LHC contour was determined using the same method as in Fig.~\ref{fig:eRportal}.  Compared with that figure, the contours here are shifted to higher masses, as the production cross section of a pair of fermionic $Q_X/e = 5/6$ particles is larger than for a pair of scalars. 

We stress that the constrained regions of parameter space in Fig.~\ref{fig:eRportal}, \ref{fig:eRportal_flipped} come from ignoring the $X$ particles entirely. If one does not have to rely on $X$ for triggering (as would be the case in this scenario where the $X$ are accompanied by energetic leptons), perhaps the subtler signatures of $|Q_X/e| = 1/6$ particles could be dug out. Even if the signal efficiency for a $\ell^+\ell^- + $ low $dE/dx$ tracks search is low and only feasible for low $M_\phi, M_X$, such a search would provide insight into otherwise unexplored parameter space.

We mention briefly that one could also consider a pair of FCPs coupled to the SM via the $L$ portal, where $e_R$ in Eq.~\eqref{eq:erportal} is swapped for the left-handed lepton doublet.  In this case, one of $\phi, X$ must carry $SU(2)$ charge, for example $\phi \sim (0,2,Y), X \sim (0,0,-(3+Y))$. Using $\phi_e$, $\phi_\nu$ to denote the upper (lower) component of $\phi$:

\begin{align}
 \mathcal L \supset \bar X(i \slashed{D} - M_X)X + |D_\mu \phi|^2 - M^2_\phi |\phi|^2 - y (e_L^\dag \, \phi_e + \nu^\dag_L \phi_\nu)\,X_R +  \rm h.c.
 \end{align}
 The phenomenology for this setup is similar to the $e_R$ version except the production cross section for a $SU(2)$ charged $\phi$ is significantly higher than $\phi$ charged under $U(1)$ alone. One component of the $\phi$ doublet will decay to $\ell + X$, while the other component decays to $\nu + X$, diluting the $\ell^+\ell^- + X\bar X$ cross-section.

Lastly, while the bounds placed at a hadron collider are independent of $y$ (as long as it is not so small that the decays of the heavier species are displaced), at a lepton collider one could be sensitive to $t$-channel production---either $\ell^+\ell^- \to \phi^* \phi$ via $t$-channel $X$, or the reverse. It would be interesting to perform a dedicated study of these signatures to help guide the development of detectors at future such lepton colliders and ensure sensitivity to fractionally-charged signatures.

\subsection{$u_R$ portal}

For our final scenario, we consider a pair of FCPs coupled to a right-handed up-type quark.  This setup is similar to the lepton portal, the main difference being that now one (or both) of the FCPs must carry color. For simplicity, and in order to emphasize ways in which a low-charge, color/$SU(2)$ singlet can be produced, we will take only one of $\phi, X$ to be colored:
\begin{align}
 \mathcal L \supset \bar X(i \slashed{D} - M_X)X + |D_\mu \phi|^2 - M^2_\phi |\phi|^2 - y \bar u P_L X\, \phi + \rm h.c.
 \end{align}
We take $\phi$ to have charges $(3,0,Y)$ while $X \sim (0,0,4-Y)$. For $\phi, X$ to be fractionally charged, we require $Y \neq 4 \ (\text{mod } 6)$. We consider interactions involving only one mass eigenstate $u_R$ for simplicity and assume the portal coupling is large enough so that all decays are prompt.

As in prior examples, the phenomenology is controlled by the mass hierarchy of the exotic states and the hypercharge $Y$. Let us focus on the scenario where $Y=3$, so that $Q_X/e = 1/6$ and assume $M_\phi > M_X$. 

In this case, and barring extremely small $y$ or degenerate $M_\phi, M_X$, $\phi$ will decay promptly to $X + j$ for portals involving $u$ or $c$ quarks and $X + t$ for a top quark portal. This leads to a final state $pp \to \phi^*\phi \to \bar X X + jj$ or $\bar XX + \bar t t$ signature. This exactly mimics a simplified supersymmetry model with only one squark (L or R) and the lightest neutralino taken to be light and all other superpartners decoupled. The only differences between our setup and the MSSM simplified model is that $X$ is fractionally charged rather than neutral, and the portal coupling $y$ is not tied to a SM gauge coupling.

The portal interaction also admits a second $pp \to \phi^*\phi$ or $pp \to \bar X X$ production mechanism $\propto y^2$ at the amplitude level via a $t$-channel $X$ or $\phi$ exchange. In the language of the MSSM, this $pp \to \phi^*\phi$ mode mimics $t$-channel squark production either through a gluino or a neutralino. If the gluino is decoupled, as in many of the scenarios in Ref.~\cite{CMS:2016eju,CMS:2019zmd}, only production through electroweakinos remains (and is negligible compared to QCD production). For portals involving up quarks and $y \sim \mathcal O(1)$, this mechanism may be more important, though the fact that $X$ are Dirac rather than Majorana means the amplitude $\sim 1/M^2_X$ (in the limit that $M_X$ is larger than the momentum exchanged) instead of $\sim 1/M_X$. The amplitude does interfere with QCD production, so the total impact on the rate will be $\mathcal O(g^2y^2)$ as well as $\mathcal O(y^4)$. Similar logic holds for $pp \to \bar XX$, which mimics MSSM gluino/neutralino production via a $t$-channel sfermion. This portal also permits $gq \to \phi^* X + h.c.$ production through an $s$-channel quark. This mode leads to a $\bar X X + j$ mono-jet signature, though its rate is $\propto y^2$. Both of these $y$-dependent production modes are negligible for charm or top portals, as the charm/top parton distribution functions are highly suppressed.  

The inclusive cross section -- which combines $pp \to \bar X X$ and $pp \to \bar X X + jj$ (or $pp \to \bar X X$ and $pp \to \bar X X + \bar tt$) -- is shown in Fig.~\ref{fig:uRportal} below, along with the ratio of the inclusive cross section to $pp \to \bar X X$ alone. On top of the cross-section plots, we show the bounds from jets plus $\slashed E_T$ bounds LHC searches and the milliQan bounds/projections. To determine the LHC bounds, we solved for the $\phi, X$ masses that saturate the quoted 95\% CL cross-section bounds quoted in~\cite{CMS:2019zmd}\footnote{Similar bounds from ATLAS can be found in~\cite{ATLAS:2020syg}}. As electroweak contributions to $\phi$ production are completely negligible, we used the quoted NLO-NLL results for squark production (assuming all other superpartners are decoupled) in Ref.~\cite{Borschensky:2014cia,Beenakker:2016lwe,Beenakker:1996ch,Kulesza:2009kq} for the $\phi$ cross section\footnote{Technically, dividing the results in~\cite{Borschensky:2014cia,Beenakker:2016lwe,Beenakker:1996ch,Kulesza:2009kq} by two as $\phi$ is only a single `chirality' squark}. The bounds we determine therefore assume that the portal coupling is small enough that we can ignore all $\phi$ production mechanisms $\propto y$. 

\begin{figure}
\centering
\includegraphics[width=0.49\textwidth]
{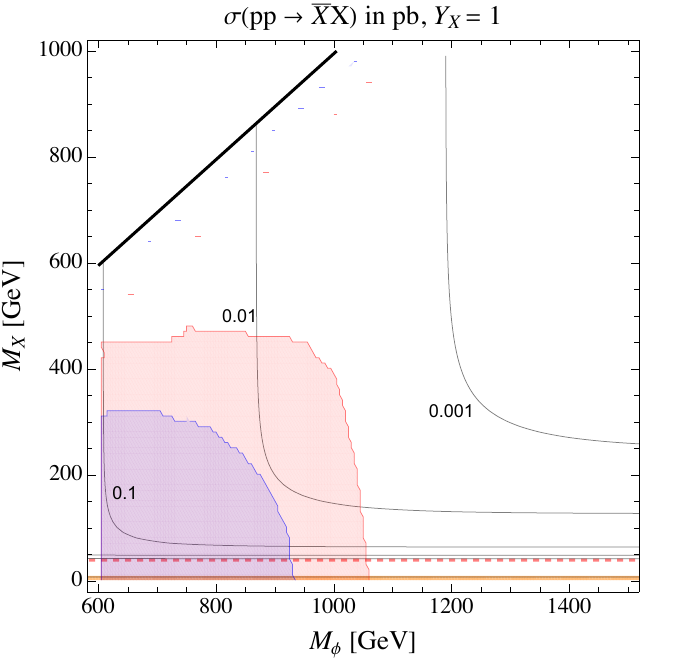}
\includegraphics[width=0.49\textwidth]
{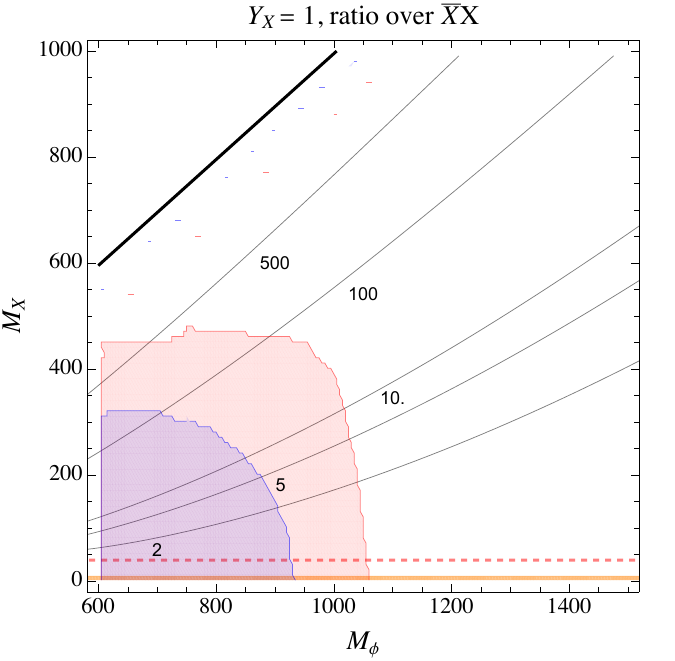}\\
\vspace{0.4cm}
\includegraphics[width=0.49\textwidth]
{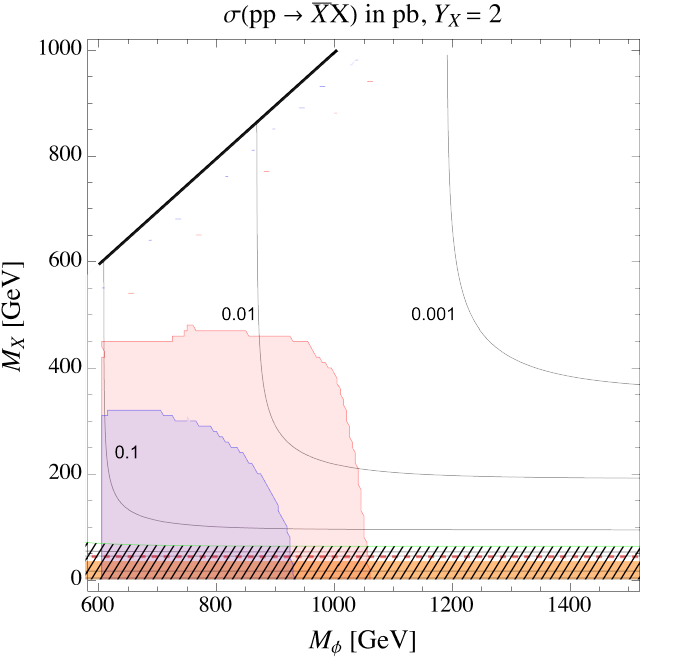}
\includegraphics[width=0.49\textwidth]
{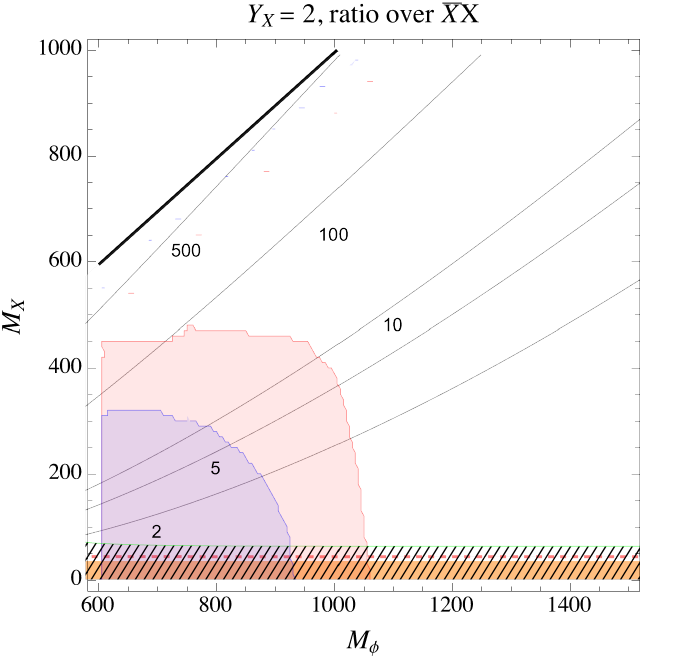}
\caption{In the left panel, the total cross-section for $pp \to \bar X X$ + $pp \to \bar X X + jj/\bar X X + \bar t t$ as a function of the FCP masses $M_\phi$, $M_X$ in the $u_R$-portal setup. We assume $M_\phi > M_X$ and assumed $y$ is small enough that production processes $\propto y$ have negligible contribution to the cross section. The shaded portion in the lower left is the excluded region from $\slashed E_T +$ jets searches at the LHC for a single flavor squark (blue for $u, c$ portals and red for top portals), assumed to decay to jet plus neutralino (or top plus neutralino, for top portal) and with all other superparticles decoupled. In the top right panel, the ratio of the cross section to $pp \to \bar X X$ and $pp \to \bar X X + jets$ to the cross section to $pp \to \bar X X$ alone. The bottom panels show the cross section, ratio, and bounds for the $u_R$ portal when $Y = 2$, such that the charge of $X$ is $1/3$. The hatched contours on the bottom panels show an estimate of the bounds from a hypothetical variation on the CMS FCP search where additional energetic charged tracks are allowed.}
\label{fig:uRportal}
\end{figure}

Inspecting Fig.~\ref{fig:uRportal}, we see that the cross-section contours are nearly vertical -- driven by the mass of the colored state $\phi$ -- until $M_X \lesssim 200\, \text{GeV}$. For lighter $M_X$, kinematics ($pp \to \bar X X$) wins over strong coupling ($pp \to \phi^*\phi$), and the contours become horizontal. 

For massless $X$ (and portals involving a $u_R, c_R$ quark), the bound on $\phi$ is $\sim 900\, \text{GeV}$. The bound of $\phi$ drops for larger $M_X$, reaching $\sim 600\, \text{GeV}$ for $M_X \sim 450\, \text{GeV}$. This trend likely continues, but is not straightforward to analyze as the search we adopt cuts off at $600\, \text{GeV}$ for the `squark'. The exact numerical bound is not particularly important, as our main point is that the bound is significantly less than in Ref.~\cite{Koren:2024xof}, where we considered $\phi \sim (3,0,3)$ without a portal interaction. These states could not decay and were therefore constrained to be $\gtrsim 1.5\, \text{TeV}$. In addition to giving $\phi$ an avenue to decay and thereby easing its constraints, the portal interaction also boosts the inclusive production of $\bar XX$. This boost, determined by the ratio relative of inclusive to exclusive $pp \to \bar X X$ production, is shown in the right hand panel of Fig.~\ref{fig:uRportal}. The boost factor can be sizable, especially for heavy $m_X$.  

The two bottom panels of Figure~\ref{fig:uRportal} show the cross sections, ratio to $pp \to \bar X X$, and the bounds if we pick $Y = 2$, so that $Q_X/e = 1/3$. The cross-section and jets plus $\slashed E_T$ contours are essentially identical.\footnote{The bounds are identical because we have assumed that $X$ with electric charge $1/3$ are invisible, as the vast majority of their track signatures are not reconstructed} The main difference compared to the top panels is that, because $Q_X/e = 1/3$, this scenario could potentially be bounded by the CMS FCP search~\cite{CMS:2024eyx}. We say potentially because these events will always contain more than two energetic tracks and would not pass the {\it current} signal selection. As in the first benchmark, we have indicated an estimate of the bounds from a modified CMS FCP search where extra tracks are permitted with the hatched contour. The bound is derived by applying the bound from \cite{CMS:2024eyx} to inclusive $pp \to \bar X X$ production.

There are two other phenomenological points for this portal that we would like to mention First, one could also consider swapping $u_R$ for a left-handed quark -- a $Q$-portal interaction (still imagining the hypercharge is chosen so that $X$ has $Q/e = 1/6$. This does not lead to any real change in the phenomenology; the $SU(2)$ charge plays little role in the production cross section, as it is totally dominated by QCD, and it does not change the detector signal, as all doublet $\phi$ decays still produce jets. 

Second, for the top quark portal (meaning that the quark involved is a right-handed top quark), there is an additional signal from four-body decay $t \rightarrow \bar X X b W$, which proceeds through an off-shell $\phi$. We estimate the branching ratio into this four-body final state as
\begin{align}
    BR(t \to \bar XXbW) \sim \frac{y^4\, m^4_t}{(16\pi^2)^2\, m^4_\phi} \sim 3.8\times 10^{-8}\, y^4 \Big(\frac{1\, \text{TeV}}{m_\phi}\Big)^4
\end{align}
for $M_X = 0$. Given that the HL-LHC ($3\, ab^{-1})$ will produce $\mathcal O(10^{10})$ top quarks, this rare decay could place interesting bounds on setups with $\mathcal O(1)$ portal couplings.

\section{Conclusions}

In this work we have studied a representative sampling of extensions of the SM by two fractionally-charged particles in such a way that there is a tree-level interaction with one of the SM species. Compared to our prior study of one-particle extensions, there is a vastly richer phenomenology made possible.

We've learned that the representations which have the best bounds when added singly may see their constraints degrade by factors of a few when destabilized by decays to a second fractionally-charged particle. Conversely, the fractionally-charged particles with the weakest bounds may see enormously more production than in the minimal scenario if they appear as decay products. Together these underscore the importance of building detectors and designing searches that can be sensitive to a broader range of final states which include FCPs. The closest improvement at hand may be to interrogate events on tape with $\slashed{E}_T$ for the presence of `low-quality' tracks from FCPs, which would not have been reconstructed with the standard track algorithm.

\section*{Acknowledgements}

We thank Julie Pages for assistance with {\tt Matchete}, David Stuart and David W. Miller for helpful conversations, and David Stuart for comments on a draft of this work.
This work was partially supported by the National Science Foundation under Grant Number PHY-2412701.

\bibliographystyle{JHEP}
\bibliography{morefractionalcharge}
\end{document}